\documentclass[a4paper]{iopart}
\usepackage{graphicx}
\begin{document}
\title[Gowdy waves as a test-bed for constraint-preserving boundary
conditions]{Gowdy waves as a test-bed for \\ constraint-preserving
boundary conditions}

\author{C. Bona and C. Bona-Casas   }

\address{Departament de Fisica, Universitat de les Illes
Balears, Palma de Mallorca, Spain.\\
Institute for Applied Computation with Community Code
(IAC$^{\,3}$)}

\ead{cbona@uib.es}

\begin{abstract}
Gowdy waves, one of the standard 'apples with apples' tests, is
proposed as a test-bed for constraint-preserving boundary
conditions in the non-linear regime. As an illustration,
energy-constraint preservation is separately tested in the Z4
framework. Both algebraic conditions, derived from energy
estimates, and derivative conditions, deduced from the
constraint-propagation system, are considered. The numerical
errors at the boundary are of the same order than those at the
interior points.
\end{abstract}

Constraint-preserving boundary conditions is a very active
research topic in Numerical
Relativity~\cite{Winicour,Rinne,Nunez}. During this decade, many
conditions have been proposed, adapted in each case to some
specific evolution formalism: Fritelli-Reula~\cite{Stewart},
Friedrich-Nagy~\cite{FN}, KST~\cite{Calabrese03,ST05},
Z4~\cite{BLPZ05},
Generalized-Harmonic~\cite{Babiuc07,Kreiss07,Winicour,Rinne}, or
BSSN~\cite{Nunez}. Cross-comparison among different evolution
formalisms has been carried out ('apples with apples'
initiative~\cite{Mexico, Mexico2}). But only periodic boundary
conditions have been considered up to now.

We endorse some recent claims (by Winicour and others) that the
cross-comparison effort should be extended to the boundaries
treatment. In this paper, we show that Gowdy waves~\cite{Gowdy71},
one of the 'apples with apples' tests, is suitable for boundary
conditions cross-comparison in the non-linear regime. As an
illustration, we test separately the energy-constraint
preservation in the Z4 framework. We compare algebraic conditions,
derived from energy estimates, with derivative conditions, deduced
from the constraint-propagation system. The resulting numerical
errors at the boundary are of the same order-of-magnitude than
those at interior points.

\section{The Gowdy waves metric}
Let us consider the Gowdy solution~\cite{Gowdy71}, which describes
a space-time containing plane polarized gravitational waves. The
line element can be written as
\begin{equation}\label{gowdy_line}
  {\rm d}s^2 = t^{-1/2}\, e^{Q/2}\,(-{\rm d}t^2 + {\rm d}z^2)
  + t\,(e^P\, {\rm d}x^2 + e^{-P}\, {\rm d}y^2)
\end{equation}
where the quantities $Q$ and $P$ are functions of $t$ and $z$
only, and periodic in $z$. The initial slice $t=t_0$ is usually
chosen so that the simulations can start with an homogeneous
lapse.

Let us now perform the following time coordinate transformation
\begin{equation}\label{gowdy_time}
  t~=~t_0\;e^{-\tau / \tau_0}\,,
\end{equation}
so that the expanding line element (\ref{gowdy_line}) is seen in
the new time coordinate $\tau$ as collapsing towards the $t=0$
singularity, which is approached only in the limit
$\tau\rightarrow\infty$. This ``singularity avoidance" property of
the $\tau$ coordinate is due to the fact that the resulting
slicing by $\tau=constant$ surfaces is harmonic \cite{BM83}. We
will run our simulations in normal coordinates, starting with a
constant lapse $~\alpha_0=1~$ at $\tau=0$ ($t=t_0$).

Standard cross-comparison tests~\cite{Mexico,Mexico2} are
currently done with periodic boundary conditions. But one gets
basically the same results by setting up algebraic boundary
conditions, which take advantage of the symmetries of the Gowdy
line element. For a rectangular grid, planar symmetry allows
trivial boundary conditions along the $x$ and $y$ directions.
Also, allowing for the fact that the $z$ dependence in
(\ref{gowdy_line}) is only through $~\cos(2 \pi z)~$, one can set
reflecting boundary conditions for the interval $~0 \leq z \leq
1~$. In this way, the Gowdy waves metric is obtained as a sort of
stationary gravitational wave in a cavity with perfectly
reflecting walls.

We can then set up a full set of algebraic boundary conditions,
which are consistent with the Gowdy line element
(\ref{gowdy_line}) for all times. This opens the door to a
selective testing procedure, where one could for instance try some
constraint-preserving condition for the longitudinal and
transverse-trace modes, while keeping the exact condition for the
transverse traceless ones. Or, as we will do below, testing just
some energy-constraint preserving boundary conditions while
dealing with all the remaining modes in an exact way.

\section{Characteristic decomposition}
We will consider here the the first-order version in normal
coordinates, as described in refs.~\cite{Z4,LNP}. For further
convenience, we will recombine the basic first-order fields
$(K_{ij},\,D_{ijk},\,A_i,\,\Theta,\,Z_i)$ in the following way:
\begin{eqnarray}
\label{Pimu a}
  \Pi_{ij} &=& K_{ij} - (\,tr K - \Theta)\, \gamma_{ij} \qquad
  V_i = \gamma^{rs}(D_{irs} - D_{ris}) - Z_k \\
\label{Pimu d}  \mu_{ijk} &=& D_{ijk} - (\gamma^{rs}D_{irs}-
V_i)\,\gamma_{jk} \qquad
  W_i = A_i - \gamma^{rs}D_{irs} + 2\,V_i
\end{eqnarray}
so that the new basis is
$(\Pi_{ij},\,\mu_{ijk},\,W_i,\,\Theta,\,V_i)$. Note that the
vector $Z_i$ can be recovered easily from this new basis as
\begin{equation}\label{Zfrom mu}
    Z_i=-{\mu^k}_{ik}.
\end{equation}

In order to compute the characteristic matrix, we will consider
the standard form of (the principal part of) the evolution system
as follows
\begin{equation}\label{fluxcons}
    \partial_t~\mathbf{u} + \alpha\,\partial_n\,\mathbf{F}^n(\mathbf{u})
    = \cdots~,
\end{equation}
where $\mathbf{u}$ stands for the array of dynamical fields and
$\mathbf{F}^n$ is the array of fluxes along the direction given by
the unit vector $\mathbf{n}$. With this choice of basic dynamical
fields, the principal part of the evolution system gets a very
simple form in the harmonic slicing case:
\begin{eqnarray}
\label{evol a}  F^n(W_i) &=& 0 \qquad  F^n(\Theta) =
V^n  \qquad  F^n(V_i) = n_i\, \Theta\\
\label{evol e}  F^n(\Pi_{ij}) &=& {\lambda}_{nij} \qquad
F^n(\mu_{kij}) = n_k \,\Pi_{ij}
\end{eqnarray}
where the index $n$ means a projection along $n_i$, and we have
noted for short
\begin{equation}\label{lambda}
    \lambda_{nij} = \mu_{nij} + n_{(i}W_{j)} - W_n\,\gamma_{ij}\,,
\end{equation}
where round brackets denote index symmetrization.

We can now identify the constraint modes, by looking at the Fluxes
of  $\Theta$ and $Z_i$ in the array (\ref{evol a}-\ref{evol e}).
It follows from (\ref{evol a}) that the energy-constraint modes
are given by the pair
\begin{equation}\label{Energy pair}
     E^\pm = \Theta~\pm~V_n
\end{equation}
with propagation speed $\pm\alpha$. Also, allowing for (\ref{Zfrom
mu},\ref{evol e}), we can easily recover the flux of $Z_i\,$:
\begin{equation}\label{evol Z}
    F^n(Z_{i}) = - {\Pi^n}_{i}
\end{equation}
so that we can identify the momentum-constraint modes with the
three pairs
\begin{equation}\label{Momentum pairs}
     M_i^\pm = \Pi_{ni}~\pm~\lambda_{nni}~,
\end{equation}
with propagation speed $\pm\alpha$. Note that, allowing for
(\ref{Pimu a}), the longitudinal component $\Pi_{nn}$ does
correspond with the transverse-trace component of the extrinsic
curvature $K_{ij}$. We give now the remaining modes: the fully
transverse ones, with propagation speed $\pm\alpha$,
\begin{equation}\label{Transverse pairs}
     T_{AB}^\pm =\Pi_{AB}~\pm~\lambda_{nAB}
\end{equation}
(the capital indices denote a projection orthogonal to $n_i$), and
the standing modes (zero propagation speed):
\begin{equation}\label{standing}
    W_i~,\qquad V_A ~,\qquad \mu_{Aij}~.
\end{equation}
Note that the standing modes (\ref{standing}), the energy modes
(\ref{Energy pair}) and the transverse momentum modes $M_A^\pm$
actually vanish for the Gowdy line element (\ref{gowdy_line}).

\section{Energy-constraint preserving boundary conditions}
In refs.~\cite{Z4,LNP}, the system above was shown to be symmetric
hyperbolic, by providing a suitable energy estimate. We can
rewrite it here as
\begin{equation}\label{estimate}
    \Pi_{ij}\Pi^{ij}+ \lambda_{kij}\lambda^{kij} + \Theta^2 + V_kV^k +
    W^kW^k
\end{equation}
This leads to the following sufficient condition for stability
\begin{equation}\label{Bterm}
    (\Pi^{ij}~\lambda_{nij} + \Theta~V_n)\mid_\Sigma ~\ge~ 0
\end{equation}
where $\Sigma$ stands for the boundary surface ($\mathbf{n}$ being
here the outward normal).

Let us consider for instance the boundary at $~z=1~$. We can
enforce there the partial set of exact (reflection) boundary
conditions $~\lambda_{nij} = 0~$, so that the requirement
(\ref{Bterm}) reduces to
\begin{equation}\label{BEterm}
    (\Theta~V_n)\mid_\Sigma ~\ge~ 0
\end{equation}
which can be used for a separate test of energy-constraint
preserving boundary conditions.

As an illustration, we will test two such conditions. The first
one is given in the form of a logical gate:
\begin{equation}\label{Thetagate}
    (\Theta~V_n)\mid_\Sigma ~< ~ 0\qquad
    \Rightarrow \qquad\Theta\mid_\Sigma ~ = 0
\end{equation}
($\Theta$-gate), so that it only acts when condition
(\ref{BEterm}) is violated. The second one is an advection
equation
\begin{equation}\label{advect}
    \partial_t~\Theta \mid_\Sigma ~ = -\alpha\, \partial_n\,\Theta
    - \eta~\Theta\,,
\end{equation}
where we have included a suitable damping term. Note that the
principal part of the constraint-preserving condition
(\ref{advect}) coincides with the 'maximal dissipation' one,
$~\partial_t~E^- = 0~$, which is not constraint-preserving in the
generic case. Condition (\ref{advect}) can be understood as a sort
of maximal dissipation condition for the energy-constraint
evolution equation
\begin{equation}
\label{Thetasub}
\partial^2_{tt}~\Theta - \alpha^2\,\triangle~\Theta = \cdots~,
\end{equation}
which follows from (the time component of) the covariant
divergence of the Z4 field equations~\cite{Z4}.

\begin{figure}[h]
\centering \scalebox{0.8}[0.8] {\includegraphics{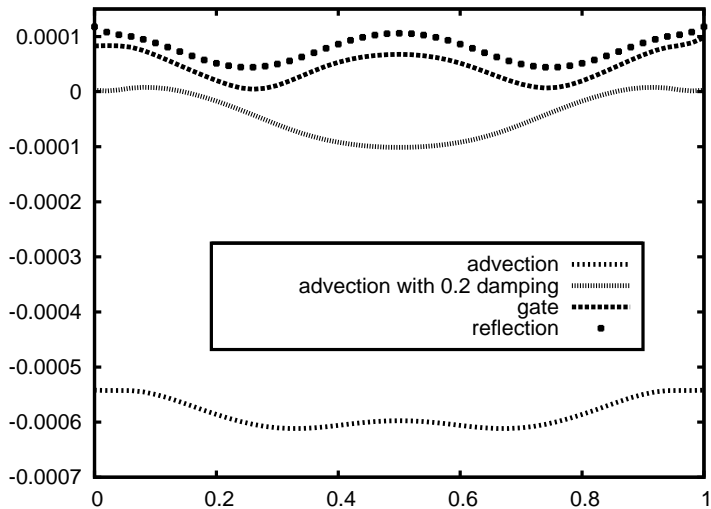}}
\caption{\label{plots} $\Theta$ profiles along the $z$ direction
for a Gowdy waves simulation ending at $\tau=100$. From bottom to
top, results for the pure advection condition (\ref{advect}),
damped advection (with $\eta=0.2$), $\Theta$-gate
(\ref{Thetagate}), and exact reflection (included here for
comparison). }
\end{figure}

We show in Fig.~\ref{plots} our results for some numerical
simulations ending at $\tau=100$. We have included in the plot the
exact (reflection) results for comparison. It is clear that the
$\Theta$-gate condition gets very close to (actually slightly
better than) the exact result in this case. The pure advection
condition (\ref{advect}) is off by half-an-order of magnitude.
However, a suitable damping term (we have taken $\eta=2$) greatly
improves this, leading in this case to even less error than the
pure reflection condition. Note that we are showing here the
$\Theta$ profiles, giving the cumulated energy-constraint
deviation. The average energy-constraint error in these
strong-field simulations is actually smaller.

\ack
This work has been jointly supported by European Union FEDER
funds and by the Spanish Ministry of Science and Education
(projects FPA2007-60220, CSD2007-00042 and ECI2007-29029-E).
C.~Bona-Casas acknowledges the support of the Spanish Ministry of
Science, under the FPU/2006-02226 fellowship.

\end{document}